\documentclass[runningheads]{llncs}

\usepackage{color}
\usepackage{graphicx}
\usepackage[bookmarks=false]{hyperref}
\usepackage{cite}
\usepackage{caption}
\usepackage{subcaption}
\usepackage{multirow}
\usepackage{rotating}
\usepackage{booktabs}
\usepackage{paralist}

\def\Snospace~{\S{}}

\newif\ifshowComments
\showCommentstrue

\showCommentsfalse

\begin{document}

\setcounter{page}{1}
\pagestyle{headings}  

\titlerunning{Field Study: Elicitation and Classification of Defects for Defect Models}
\title{A Field Study on the Elicitation and Classification of Defects for Defect Models}
\author{D. Holling \and D. M\'{e}ndez Fern\'{a}ndez \and A. Pretschner}
\institute{Technische Universit\"at M\"unchen, Germany\\ \email{\{holling, mendezfe, pretschn\}@cs.tum.edu}}


\newcommand{\changed}[1]{{\textbf{\color{red}{CHANGED: #1}}}}
\newcommand{\fixme}[1]{{\color{red}{\textbf{FIX: #1}}}}

\maketitle

\begin{abstract}
\textbf{Background:} Defect models capture faults and methods to provoke failures. To integrate such defect models into existing quality assurance processes, we developed a defect model lifecycle framework, in which the elicitation and classification of context-specific defects forms a crucial step. Although we could gather first insights from its practical application, we still have little knowledge about its benefits and limitations. \textbf{Objective:} We aim at qualitatively analyzing the context-specific elicitation and classification of defects to explore the suitability of our approach for practical application.  \textbf{Method:} We apply case study research in multiple contexts and analyze (1) what kind of defects we can elicit and the degree to which the defects matter to a context only, (2)  the extent to which it leads to results useful enough for describing and operationalizing defect models, and (3) if there is a perceived additional immediate benefit from a practitioner's perspective. \textbf{Results:} Our results strengthen our confidence on the suitability of our approach to elicit defects that are context-specific as well as context-independent. 
\textbf{Conclusions:} We conclude so far that our approach is suitable to provide a blueprint on how to elicit and classify defects for specific contexts to be used for the improvement of quality assurance techniques.

\end{abstract}
\vspace{-3em}
\section{Introduction}
\vspace{-1em}
\label{sec:introduction}
Defect models capture faults and methods to provoke failures~\cite{Pretschner:2013, pretschner:defectModels:2015} and describe them formally. By operationalization, formal defect models can be used as a basis to create (semi-) automatic defect detection tools. On the one hand, such tools include (semi-)automatic test case generators to detect smells and gather evidence for described faults or execute certain test strategies for methods to provoke failures. On the other hand, they also include checklist generators or reading technique organizers to detect defects in non-executable, and therefore, non-testable artifacts. Since operationalizations directly target the described defects, they support systematic fault-based testing and yield good test cases \cite{Pretschner:2013}.

In our definition, a~\emph{defect} is an umbrella term including all faults, errors, failures, bugs, and mistakes made when designing or implementing a system. Similar to the notion of quality in general, which constitutes a multifaceted topic with different views and interpretations~\cite{1984_garvind_product_quality, KP96}, defects and especially their relevance, too, are something relative to their context. That is, a defect that might be critical to one project might be without relevance to the next. The systematic integration of (domain-specific) defect detection and prevention mechanisms into the quality assurance (QA) of particular socio-economic contexts, e.g. a company or a business unit is therefore crucial. 

To integrate defect models into existing quality assurance processes, we developed and previously published a defect model lifecycle framework~\cite{Holling:2014:ICST}.
This framework can be embedded into established quality improvement lifecycle models and provides a blueprint of steps and artifacts to plan, employ, and control defect models. A planning step comprehends the elicitation and classification of defects (or defect classes respectively) in order to later on describe their defect models formally and operationalize them in an employment step. 

The construction of formalized defect models and their operationalization in tools is very expensive. 
Eliciting and classifying the ``relevant'' defects for specific contexts is thus crucial for taking the context-specific decision whether to invest this effort. 
Defect elicitation and classification methods hence needed to be comprehensive and to allow for frequency, and possibly severity, assessments. Based on these assessments, the effectiveness of defect models can be anticipated and the investment decision can be rationalized. 
%
%


The subject of this paper is our Defect ELIcitation and CLAssification approach (DELICLA). We aim at understanding the effectiveness of DELICLA by means of qualitative methods with particular focus on interviews for the data collection and Grounded Theory for the analysis. One reason for relying on Grounded Theory coding principles is the categorization as well as the possible elaboration of cause-effect relations for defects. Once the defects are identified, they are integrated in a taxonomy: technical or process-related. The qualitative nature makes the approach agnostic to specific contexts/domains while, at the same time, always yielding context-specific results. By relying on an adaptable defect taxonomy, we follow the baseline of Card \cite{Card:2005} and Kalinowski et al. \cite{Kalinowski:2010}, who note that it is beneficial to ``tailor it to our [...] specific needs''.



\textbf{Problem Statement.}  Although we had gathered first insights from applying DELICLA in practice, we have yet little knowledge about its appropriateness to (1) elicit and classify defects in specific contexts of different application domains; (2) the extent to which it leads to results useful enough for describing and operationalizing defect models; and (3) if there are immediate additional benefits as perceived by practitioners. These problems are reflected by the research questions in Section \ref{sec:evaluation}.

\textbf{Contribution.} We contribute a field study where we apply our DELICLA approach to four cases provided by different companies. In each case, we conduct a case study to elicit and classify context-specific defect classes. The goal of our study is to get insights into advantages and limitations of our approach; this knowledge supports us in its further development. 

Researchers as well as practitioners can directly apply our approach and the resulting defect taxonomies which include common and recurring defects. In addition to the defect-based perspective, we explicitly provide tacit knowledge about defects useful to organizations in order to advance in organizational learning~\cite{Schneider:2009}. By evaluating our approach in different contexts, we lay the foundation for its adoption in research and practice.

\section{Related Work}
\vspace{-1em}
\label{sec:relatedwork}
In the classification step of our approach, we provide a basic defect taxonomy / classification. Efforts to create a standardized defect classification for the collection of empirical knowledge have been made in the past \cite{Wagner:2008}. However, there has not yet been a general agreement as defects may be very specific to a context, domain or artifact. This leads to a plethora of taxonomies and classifications techniques in literature and practice. In the area of taxonomies, Beizer \cite{beizer} provides a well-known example for a taxonomy of defects. Avizienis et. al. \cite{Avizienis:2004, Avizienis:2004:2} provide a three-dimensional taxonomy based on the type of defect, the affected attribute and the means by which their aim of dependability and security was attained. IEEE standard 1044 provides a basic taxonomy of defects and attributes that should be recorded along with the defects. Orthogonal Defect Classification (ODC) \cite{ODC} is a defect classification technique using multiple classification dimensions. These dimensions span over cause and effect of a defect enabling the analysis of its root cause. Thus, defect trends and their root causes can be measured in-process. Apart from these general classification approaches, there are approaches specifically targeting non-functional software attributes such as security \cite{Aslam:1996,Landwehr:1994:1,Landwehr:1994:2} or, based upon ODC, maintenance \cite{Ma:2003, Ma:2007}. Leszac et al. \cite{Leszak:2002} even derive their classification aiming to improve multiple attributes (i.e reliability and maintainability). Our approach, presented next, deliberately chooses to employ a minimalistic/basic defect taxonomy to stay flexible for seamless adaptation to specific contexts and domains. This lightweight taxonomy enables the approach to be in tune with the expectations/prerequisites of our project partners (see RQ3 in Sect. \ref{sec:evaluation}). In contrast to ODC, we are not generalizing our taxonomy to be ``independent of the specifics of a product or organization'' \cite{ODC}, but rather require adaptability to context. In addition, we do not aim to capture the effects of defects (other than the severity where possible) as it is not required for the elicitation and classification of defects for defect models. However, our taxonomy can be mapped to ODC's cause measures by (1) refining the categories of technical and process-related defects into defect types and (2) using the associated tasks of the role of the interview partner as defect trigger. The severity can directly be taken in ODC's effect measures, but other required measures such as impact areas, ``reliability growth, defect density, etc.'' \cite{ODC} must be elicited in addition. 


\vspace{-1em}
\section{DELICLA: Eliciting and Classifying Defects}
\label{sec:Process}
\vspace{-1em}
A first decision in the design of DELICLA was to use a qualitative approach for defect elicitation and classification. The central aspect of our approach is further its inductive nature where the focus is on generating theories rather than testing given ones. That is, the approach makes no a-priori assumptions about which defects might be relevant in a specific context, yet our hypothesis is that common and recurring defects exist in the context. In addition, we rely on circularity yielding further defects, if the approach is repeatedly used in the same or similar contexts. 

There exists a multitude of techniques employable in qualitative explorative approaches with the ability to take a defect-based viewpoint. These established techniques have been explored with respect to three goals: (1) cost-effectiveness in their application, (2) comprehensiveness in the obtained results, and (3) ability to establish a trust relationship during the data collection.

Trust is important because humans generally are reluctant to disclose potential problems in individual project environments~\cite{Gubrium:2001, Kalinowski:2010}. The assessed techniques include techniques for document analyses, interview research, participant observation, and creativity techniques such as brainstorming.

Due to their comprehensiveness and the possibility to establish a trust relationship~\cite{Gubrium:2001, Hove:2005}, personal interviews were chosen as technique in the DELICLA approach. This allows to fully explore the participants' perspectives in their context while adopting their vocabulary. Using this technique in a semi-structured form yields the ability to guide the interview~\cite{Gubrium:2001} along predefined questions without interrupting their flow of words. 

For the analysis of the collected data, we employed Grounded Theory~\cite{Glaser:1967} and code the answers as described by Charmaz~\cite{Charmaz:2006}. In a manual coding step, we code all mentioned defects including their cause and effect. These codes are then organized in a hierarchy representing a defect taxonomy. Following this form of open coding, we apply axial coding to the results to explicitly capture relationships between defects as well as possible causes and effects. In some cases, we apply selective coding to capture possible causalities between defects. Our DELICLA approach consists of the three steps explained next: (1) \emph{Preparation}, (2) \emph{Execution}, and (3) \emph{Analysis}.

%

\textbf{Preparation.}
The first activity in the preparation step is to create a pool of potential interview candidates (i.e. the participants). Candidates are identified with the project partner by focusing on their projects or domains of expertise. The selection of interview candidates is performed by the interviewer or project partner yielding a variation point. In case the interviewer is able to select the candidates, the context of the study (e.g. the projects and teams focused on) and the expenditure of time for the project partner must be exactly defined. Key aspects to consider before selecting any interview partners are the organizational chart and the assessment of their potential contributions by their managers. The order of interviews was from best to least contributing according to the executives' opinions; and lowest to highest branch in the organizational charts \cite{Gubrium:2001}. When interviewing the best performing, the interviewer is able to assess the maximum capabilities of team or project members thereby gaining a perspective of what can be achieved. Subsequent to interviewing executives on higher branches, defects collected in lower branches can be discussed and used to devise first indications towards future measures. Thus, even if the project partner selects the interview partners, the interviewer should be able to get an overview using an organizational chart and set the order of the interview partners.


After the interview partners have been selected, they are informed about the upcoming interview and their required preparation. An interview preparation sheet is given to them detailing the purpose of the interviews and the questions to be prepared. In our studies, we used the open questions seen in Tab.~\ref{table:interview_questions} for preparation similar to those presented by Charmaz~\cite{Charmaz:2006}. An extension point are additional questions. Depending of the context, questions such as ``How meticulously is the SCRUM methodology followed?'' may be added. When informing the interview partners, the responsibles on the project partner's side must also be named for potential inquiries of interview partners about internal procedures. Interviews are not part of the everyday working life of the interview partners and may cause feelings of nervousness to anxiousness. To mitigate these feelings, the description of the purpose of the interviews is very detailed and emphasizes the defect-based view on tools, processes and people in defect models for quality assurance.
\vspace{-4mm}
\begin{table}[htb]
\scriptsize
\centering
\caption{Instrument used for the interview preparation sheet.} 
\label{table:interview_questions}
\begin{tabular}{ll}
\hline \hline
ID & Question \\
\hline 
Q1 & What are the classical faults in the software you review/test? \\
Q2 &What does frequently/always go wrong? With which stakeholder? \\
Q3 & What was the ``worst'' fault you have ever seen? Which one had the ``worst'' consequences? \\
Q4 & Which faults are the most difficult ones to spot/remove? \\
Q5 & What faults were you unaware of before working in your context? \\
Q6 & What faults do you find most trivial/annoying? \\
Q7 & What faults do engineers new to your area make? \\
\hline
\end{tabular}
\end{table}

Each interview requires 30 minutes for the preparation by the interview partner and 30 minutes for the actual interview; usually a negligible amount of time. This lets interview partners prepare so that they ``can be prepared to speak directly to the issues of interest''~\cite{Gubrium:2001}. When planning the concrete times for the interviews, every two interviews include a 30 minute break at the end. In case any interview takes more time than expected, this break is used to prevent the accumulation of delay for the following interviews. 

The interviewer must also prepare w.r.t. the processes and tools employed by the interview partners and their roles at the project partner. To establish the trust relationship, an address of reassurance is prepared to be given before the interview. In addition, the room is small and any distractions are removed. All technology used during the interviews is tested beforehand and interviews are recorded as suggested by Warren \cite{Gubrium:2001}. 

\textbf{Execution.}
With trust and comprehensiveness of results our main objectives, we follow the basic principles of interview research: At the beginning of the interview, the interview partner and interviewer agree on a first name basis. This basis takes down psychological walls and is a key enabler of an open discussion later in the interview. When sitting down, the interviewer never faces the interview partner as it creates the sense of an interrogation~\cite{Gubrium:2001}. The interview starts with a short introduction consisting of a description of the survey, its goals and the reasons for personal interviews. This introduction aims to mitigate any fears and allows the interview partners to get used to the interview situation. The interviewer can display knowledge and emotional intelligence at this point by stating that elicited defects will be used rather than judged for example. At the end of the introduction the way of documenting the interview results is agreed upon. There, a trade-off might be necessary between recording the interview results and manually documenting them; we experienced recordings to threaten the validity by potentially influencing the behavior of the participants while manual documentation might be prone to bias. In any case, the anonymity of the analysis is guaranteed before the interview.

Following the introduction is the description of the context by the interview partners. This includes the tasks, activities and processes they are involved in. This part of the interview is individual and helps the interviewer later in the classification of the discussed defects. Questions such as ``What are your inputs and outputs?'' help the interview partners to express their role, constraints, tasks and results toward the interviewer. The semi-structured approach of the interview helps the interviewer in this part as it allows for inquiries by the interviewers in case of unfamiliar terms and concepts. This part is not described on the questionnaire as interview partners are typically able to elaborate their work context. This also helps them to get into a flow of words as ``at a basic level, people like to talk about themselves''~\cite{Gubrium:2001}. 

The core part of the interview is the discussion of defects including their description, frequency and severity. Also the resulting failures and possible detection and/or prevention techniques are discussed. Again, this part is individual, but is guided by the questions on the interview preparation sheet. This guidance exploits the order in the heads of the interview partners as they likely prepared the questions in the order they were on the preparation sheet. At the end of the interview, an agreement of future contact has to be reached. 

In general, it is the interviewer's job to keep up an objective atmosphere and tone. It is hard for humans to admit defects and discuss them, but it is in fact the decisive point of the presented approach. Thus, the interviewer must cater to the interview partner using emotional intelligence. Additionally, ``whatever the training and intentions of the interviewer, the social interaction of the qualitative interview may unfold in unexpected ways.''~\cite{Gubrium:2001}.

\textbf{Analysis.}
The analysis of the interviews is used for the classifications of the collected defects. Defects interesting for the description and operationalization of defect models are common and recurring defects and defects with a high severity. To perform the classification and go from defects to defect classes, the first step of Grounded Theory~\cite{Glaser:1967} is employed. In that step, the recordings are coded in chronological order whereas the codes are iteratively abstracted to categories eventually leading to a basic defect taxonomy. Codes may also include contexts, roles and distinctions of the employed quality assurance process. This helps the interviewer to capture ``what is happening in the data'' \cite{Gubrium:2001}. After coding, all excerpts of the recordings are grouped by code and reheard to focus on one particular defect and its context, origination and consequences.

In the classification, the basic taxonomy of defects contains two basic families of defects: technical and process-related defects. Process-related defects concern all methodological, organizational and process aspects (as defined by the defect causal analysis \cite{Kalinowski:2008}) and contain defects causing technical defects. Technical defects are directly attributable to the product and are detectable by measures of quality assurance. These two families of defects yield extension points. An exemplary extension could be tool-related defects or defects rooted in the behavior of humans. These can be added dynamically and defects may belong to multiple classes depending on the context. Recall that, we deliberately chose to ``tailor [the taxonomy] to our [...] specific needs''\cite{Card:2005, Kalinowski:2010} to stay flexible for seamless adaptation to specific contexts and domains w.r.t. the creation of defect models.

After the analysis, we created a report summarizing the results to the project partners and used it as basis for discussion in a concluding workshop. In this workshop, we presented the results and the discussion yielded a last validation of the results w.r.t. the expectations of the project partners. Afterwards, eligible defects for the creation and operationalization of defect models were discussed and selected constituting a last contact with the project partners to potentially initiate the development of tools based on the defect models.
\vspace{-1em}
\section{Field Study Design}
\label{sec:evaluation}
\vspace{-1em}
We conducted our field study by relying in total on four cases. In each case, we follow the same study design. In the following, we report on the design which we organize according to Runeson et al.~\cite{runeson09}. 

\vspace{-1.5em}
\subsection{Research Questions}

The goal is to investigate the advantages and limitations in the elicitation and classification of defects for defect models using our DELICLA approach described in section \ref{sec:Process}. To this end, we formulate three research questions.

\textbf{RQ 1 (Suitability):} What (kind of) defects can be elicited with the approach; what is the degree of sensitivity to their context; and how comprehensive is the approach?

The core idea behind the approach is to elicit and classify common and recurring / severe defects independent of the context it is used in while preserving the context-dependent usefulness to adapt QA techniques to those defects. Hence, our first research question targets the adaptability of the approach to different employment contexts and its ability to always elicit and classify defects relevant to quality assurance independent of context. In particular, it should not be affected by changes of domains (information and cyber-physical), test / quality assurance levels (review, inspection, unit, integration and system test) and project partner. Finally, we rate a defect as context-independent if we find a relation to existing evidence in a given baseline. This means, if we find a study that indicates to the same defect in a different context, we may assume that the defect is context independent.

\textbf{RQ 2 (Operationalizability):} Can the results of the approach be used for the description and operationalization of defect models? 

The classification and elicitation of defects for defect models aims at their later description and operationalization. Thus, the results of the approach must yield a basis for decision-making to make the effort to describe and operationalize the respective defect models and yield starting points for their description and operationalization. This research question therefore aims at analyzing whether the basis of decision-making and starting points are retrievable by the approach, thereby manifesting a direct usefulness to project partners. We do not have a clear oracle to answer this question.
To answer the research question, we will therefore point to indicators for successful description and operationalization of defect models based on our approach. 

\textbf{RQ 3 (Indirect short-term benefit):} Besides potential defect models and their operationalization, how valuable are our results to the project partners? 

When our approach has been applied, project partners are given a final report to inform them about the results. This report contains all elicited and classified defects as well as possible proposals for action. In addition to the value for defect models. 
This research question targets the usefulness of the report in the eyes of those project partners considering the time invested on the project partner's side, thereby manifesting the indirect benefit of the approach. Again, we do not have an oracle. However, the quality of the results w.r.t. sufficiency and the cost-effectiveness can be rated by the project partners based on subjective expert judgement and feedback gathered during a concluding workshop. 
\vspace{-1.5em}
\subsection{Case and Subject Selection}

We apply our process for the elicitation and classification of defect models to four software development projects of different industry project partners. We do not change our process throughout the field study to gain comparable results, although this affects internal validity. The four projects were chosen on an opportunistic basis. As we required real-world development projects and project managers / members to agree, the process was performed when possible. However, the chosen cases are suitable to answer our research questions if the selected projects are distributed across different companies working in different application domains.

\vspace{-1.5em}
\subsection{Data Collection and Analysis Procedures}
\vspace{-0.8em}
To collect and analyze the data, we use our DELICLA approach for the elicitation and classification of defects as described in Sect.~\ref{sec:Process}.

To answer \emph{RQ 1}, we list the top 14 defects (i.e. all defects mentioned in at least two interviews within the same context) we elicited and classified and evaluate their commonality in contrast with their context sensitivity. That is, for each defect, we analyze whether it is context-dependent or context-independent if we find a relation to existing evidence. As a baseline, we use the defects reported by Kalinowski et al.~\cite{Kalinowski:2010} and Leszak et al.~\cite{Leszak:2002}.

We also quantify the number of defects elicited and give an assessment as to if the interviews allow for a comprehensive defect-based perspective on projects or organizations. There is no clear agreement on a sufficient number of interviews in general, but indicators toward sufficient numbers may be given \cite{Baker:2012}. For our cases, we agree on the sufficiency of the number of interviews when we observe a saturation in the answers given, i.e. when no new defects arise. Saturation is taken as a sign of comprehensiveness. 

To answer \emph{RQ 2}, we list indicators of tools and methods created from defects classified and elicited with our approach. These tools and methods may not have a fully formulated formal defect model description, but are able to demonstrate whether (and how) results may be operationalized.

To answer \emph{RQ 3}, we describe indicators of the quality of the results and the involved costs by gathering expert feedback from project partners after performing our approach. This feedback is a direct external grading of our approach by industry experts and yields an assessment of its cost-effectiveness.

\vspace{-1.5em}
\section{Case Study Results}
\vspace{-1em}
We performed the case study in four different industry projects (settings) with different industry partners. For reasons of non-disclosure agreements, we cannot give detailed information on project-specifics and the particularities of context-specific defects. However, we can state their domain, the number of interviews conducted and the classes of defects. 

\begin{figure}[ht]
\centering
\begin{subfigure}{0.49\textwidth}
	\includegraphics[width=\textwidth,height=4cm]{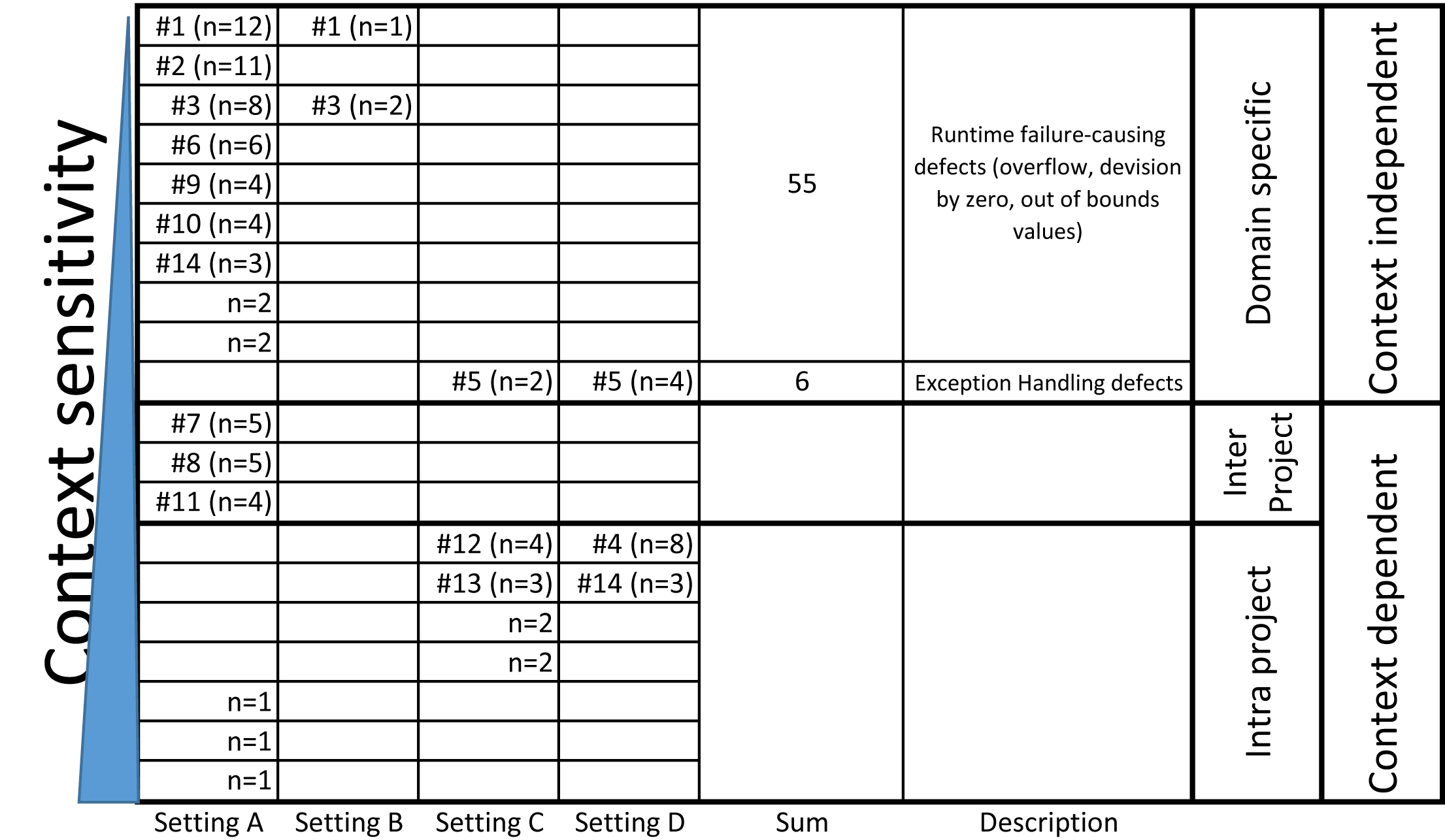}
	\caption{Technical defects}
		\label{fig:technicalDefects}
\end{subfigure}
\begin{subfigure}{0.49\textwidth}
	\includegraphics[width=\textwidth,height=4cm]{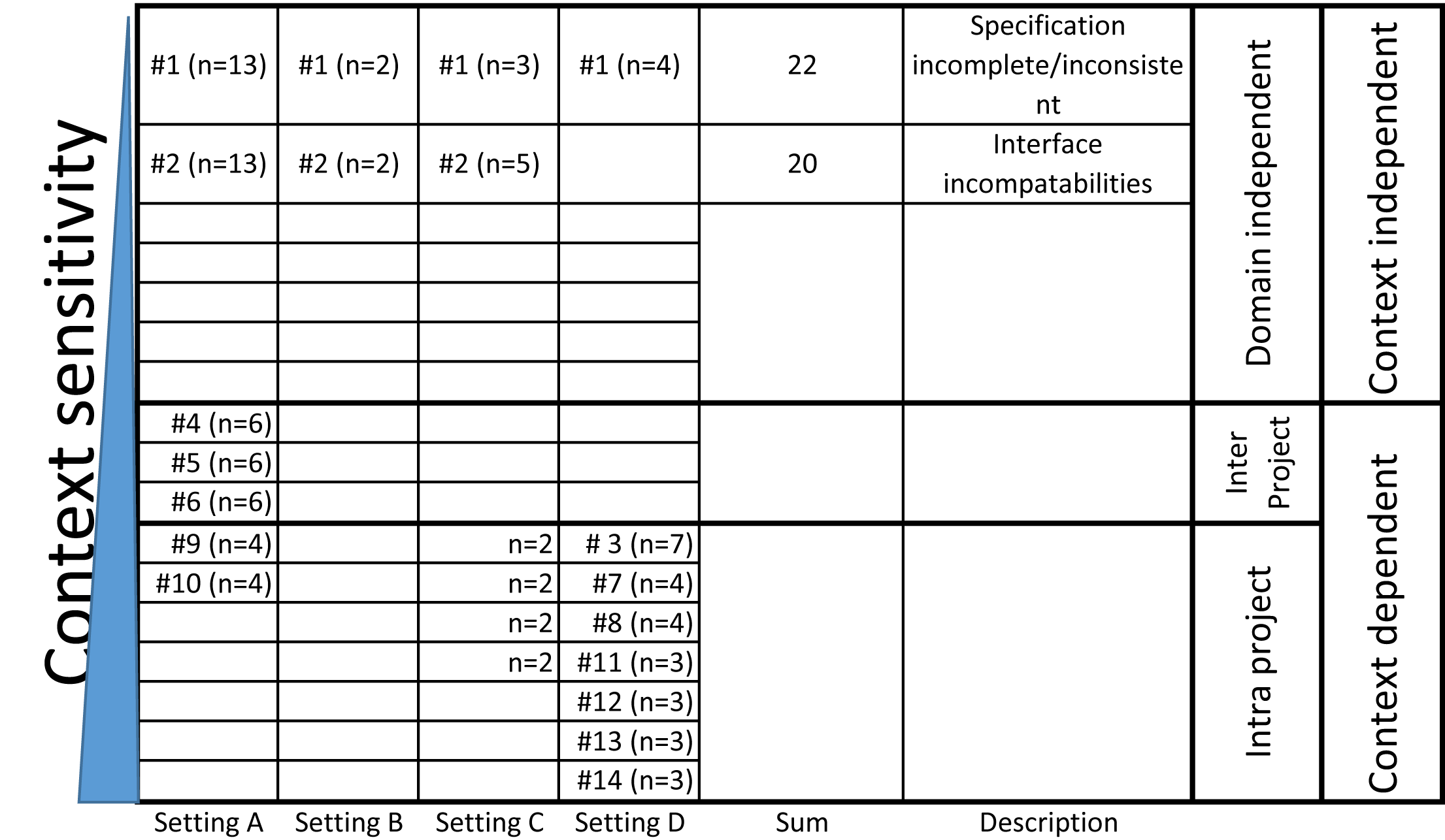}
	\caption{Process-related defects}
        \label{fig:processDefects}
\end{subfigure}
\caption{Defects with applicable ID (\#, if number in top 14) and frequency (n) in the classification by setting, domain and context specificity}
\label{fig:defects}
\end{figure}

The top 14 defects independent of their setting are shown in Tab.~\ref{tab:top15Defects}. The settings and their respectively elicited and classified defects are shown in Fig.~\ref{fig:defects}. They are grouped by our basic taxonomy defined in Sect. \ref{sec:Process} into technical (Fig.~\ref{fig:technicalDefects}) and process-related defects (Fig.~\ref{fig:processDefects}) and ordered each according to their context-sensitivity. Interestingly, we have found existing evidence for defects identified as context dependent as the existing evidence provided an extensive, and thereby, fitting defect description.


\begin{table}[htbp]
\scriptsize
  \centering
  \caption{Top 14 defects by frequency \\(at least mentioned twice in the same context (n\textgreater 2) from 43 interviews)}
    \begin{tabular}{p{0.1cm}p{0.1cm}|p{2cm}|l|p{4.5cm}|p{4.5cm}|}
    \toprule
    \textbf{} & \multicolumn{1}{c}{\textbf{\#}} & \textbf{Name} & \textbf{Ex. Ev.} & \textbf{Description} & \textbf{Mentioned Consequences} \\
    \midrule
    \multicolumn{1}{c}{\multirow{15}[2]{*}{\begin{sideways}Top 15 Technical Defects\end{sideways}}} & \multicolumn{1}{l}{1} & Signal Range & \cite{Leszak:2002} & Ranges of signals were not as described in the specification & Undefined / unspecified behaviour of connected systems  \\
    \multicolumn{1}{c}{} & \multicolumn{1}{l}{2} & Scaling & \cite{Leszak:2002} & Fixed-point values were scaled incorrectly for their specified range & Possible under-/overflows and/or system outputs differ from specification \\
    \multicolumn{1}{c}{} & \multicolumn{1}{l}{3} & Wrong initial value & \cite{Leszak:2002} & The initial values of the system were not set or set incorrectly & Initial system outputs differ from specification \\
    \multicolumn{1}{c}{} & \multicolumn{1}{l}{4} & Data dependencies & \cite{Kalinowski:2010} \cite{Leszak:2002} & Data dependencies were unclear & When changing data formats, not all locations of the data formats were updated \\
		    \multicolumn{1}{c}{} & \multicolumn{1}{l}{5} & Exception Handling & \cite{Leszak:2002} & Exception handling was either untested or not implemented as specified & Execution of exception handling routines lead to system failure \\
    \multicolumn{1}{c}{} & \multicolumn{1}{l}{6} & Dead code due to safeguards & \cite{Leszak:2002} & Dead superfluous safeguards were implemented & Degraded system performance and/or real-time requirements not met \\
    \multicolumn{1}{c}{} & \multicolumn{1}{l}{7} & Linkage of Components & \cite{Leszak:2002} & Interfaces of components were not connected as specified & System outputs differ from specification \\
    \multicolumn{1}{c}{} & \multicolumn{1}{l}{8} & Variable re-use & \cite{Leszak:2002} & Mandatory re-use of variables and developers assumed incorrect current values & System outputs differ from specification \\
    \multicolumn{1}{c}{} & \multicolumn{1}{l}{9} & Different base & \cite{Leszak:2002} & Calculations switched base (10 to 2 and vice versa) & System outputs differ from specification \\
		\multicolumn{1}{c}{} & \multicolumn{1}{l}{10} & State chart defects & \cite{Leszak:2002} & Defects related state charts & System outputs differ from specification \\
    \multicolumn{1}{c}{} & \multicolumn{1}{l}{11} & Transposed characters & \cite{Leszak:2002} & Characters in user interfaces and framework configurations were transposed & Mapping of code to user interface does not work \\
    \multicolumn{1}{c}{} & \multicolumn{1}{l}{12} & Web browser incompatibilities & & Web browsers had different interpretations of JavaScript and HTML & Browser-dependent rendering of web pages \\
    \multicolumn{1}{c}{} & \multicolumn{1}{l}{13} & Validation of input & & Inputs were either not or not validated according to specification & Ability to input arbitrary or malicious data \\
    \multicolumn{1}{c}{} & \multicolumn{1}{l}{14} & Concurrency & & Concurrency measures were not used as specified & Deadlocks and atomicity violations \\
		\toprule
    \multicolumn{1}{c}{\multirow{15}[1]{*}{\begin{sideways}Top 14 Process-related Defects\end{sideways}}} & \multicolumn{1}{l}{1} & Specification incomplete/inconsistent & \cite{Kalinowski:2010} \cite{Leszak:2002} & The specification was either inconsistent, lacking information or inexistent & Thorough verification of implementation impossible and testing deferred \\
    \multicolumn{1}{c}{} & \multicolumn{1}{l}{2} & Interface incompatibilities & \cite{Leszak:2002} & Agreed interfaces of components were not designed as discussed / specified & Rework for interfaces required after deadline for implementation \\
    \multicolumn{1}{c}{} & \multicolumn{1}{l}{3} & Missing domain knowledge & \cite{Kalinowski:2010} & Engineers lacked the concrete domain knowledge to implement a requirement & Delivery delayed and project time exceeded \\
    \multicolumn{1}{c}{} & \multicolumn{1}{l}{4} & Cloning & & Engineers used cloning as a way to add functionality to systems & Cloned parts provide functionality not required by the system in development \\
    \multicolumn{1}{c}{} & \multicolumn{1}{l}{5} & Static Analysis runtime & \cite{Leszak:2002} & Static analysis of the implementation was started too late in the process & Delivery delayed and project time exceeded \\
    \multicolumn{1}{c}{} & \multicolumn{1}{l}{6} & Quality assurance deemed unnecessary & & Engineers did not see the necessity for quality assurance & Review / Testing not performed according to specified process \\
    \multicolumn{1}{c}{} & \multicolumn{1}{l}{7} & Late involvement of users & \cite{Leszak:2002} & Users were involved late or not at all in a SCRUM-based process & Requirements not according to user problem statement \\
    \multicolumn{1}{c}{} & \multicolumn{1}{l}{8} & Misestimating of costs & \cite{Leszak:2002} & Inability to estimate cost for requirements in a SCRUM-based process & Delivery delayed and project time exceeded \\
    \multicolumn{1}{c}{} & \multicolumn{1}{l}{9} & Call order dependencies & \cite{Leszak:2002} & Call orders were switched without informing engineers & Extra testing effort required with difficult fault localization \\
    \multicolumn{1}{c}{} & \multicolumn{1}{l}{10} & Misunderstood instructions & & New engineers did not understand given documentation & Delivery delayed and project time exceeded \\
    \multicolumn{1}{c}{} & \multicolumn{1}{l}{11} & Distributed development & \cite{Leszak:2002} & Development and test team were at different locations & Communication deficiencies yielded untested components with runtime failures \\
    \multicolumn{1}{c}{} & \multicolumn{1}{l}{12} & Insufficient test environment & \cite{Leszak:2002} & The test environment did not contain all components to be tested & Some defects could only be detected in production environment \\
    \multicolumn{1}{c}{} & \multicolumn{1}{l}{13} & Development by single person & & A single person was developing a large part of the system & Incomprehensible implementation of components \\
    \multicolumn{1}{c}{} & \multicolumn{1}{l}{14} & Overloaded employees & & Engineers were overwhelmed with the amount of work requested from them & Careless misateks due to stress \\
    \bottomrule
    \end{tabular}%
  \label{tab:top15Defects}
\end{table}

\vspace{-4em}
\subsection{Case Description}
\vspace{-0.8em}

Setting A is a medium size cyber-physical software supplier. 24 subjects were interviewed with the aim to draw a organization-wide picture of common and recurring (mentioned in at least 2 interviews) defects. These systems primarily targeted the automotive domain, but also were in the domain of aerospace, railway and medical. The predominant development process was the V-model.

Setting B is a department of a large German car manufacturer. 3 subjects were interviewed as to try out the approach and enable a first glance at a defect-based perspective in this department using the V-model as development process. Note that, this low number of interviews is discusses in threats to validity.

Setting C is a project of medium size in an information system developing company. 6 subjects were interviewed to give the company an introduction to the approach. The interviews were performed in a large scale website front end project developed using the SCRUM methodology. 

Setting D was an information system project of a railway company. 10 subjects were interviewed to show process deficiencies and give a defect-based perspective on currently employed development and quality assurance measures. The project was a graphical rail monitoring application project developed using the SCRUM methodology. 

%

\vspace{-1.5em}
\subsection{Subject Description}
\vspace{-0.8em}
As described in the subject selection, we chose our project partners and projects in an opportunistic manner. All interview participants had an engineering or computer science background and at least one year of experience. The first author applied the approach in the case studies.

In setting A, the majority of participants had a background in mechanical or electrical engineering and developed system using Matlab Simulink with either automatic code generation or using Matlab Simulink models as specification for their manually implemented systems.

In setting B, the interview partners were developing and/or testing the functional software for an electronic control unit developed in C++ and integrated into AUTOSAR.

In setting C, the interview partners included a broad selection of roles including architects, developers, testers, test managers, and scrum masters.

In setting D, the interview partners were from several different teams defined in SCRUM to also gain a comprehensive view on synergy effects and defects missed by their managers.

\textbf{RQ 1: Suitability.}
In all studies performed, the results always yielded technical and process-related defects. The top 14 defects of each category are shown in Tab.~\ref{tab:top15Defects}. For each defect, we additionally show whether we could find a relation to existing evidence (see column 4 in Tab.~\ref{tab:top15Defects}). Figure \ref{fig:defects} further illustrates each defect (via its identifier provided in Tab.~\ref{tab:top15Defects}) in relation to its degree of sensitivity to the context. Remember from the introduction that ``context'' here refers to a specific company or business unit of a company.

In setting A, the interviews revealed 15 technical and 7 process-related defects. The technical defects were mainly run-time failures such as overflow due to the abstraction from the underlying computational model in Matlab Simulink. These failures were caused by wrong signal ranges of units, wrong scaling of fixed-point types and wrong initial values. The process-related defects were interface incompatibilities and incomplete / incorrect specifications. We performed 24 interviews in total. However, the top most common and recurring faults were named in 12, 11 and 8 interviews respectively. Since this was a cross-project company wide survey, the diversity of developers and testers interviewed introduced differences in the defects common and recurring in their respective fields. Baker and Edwards \cite{Baker:2012} hint at 12 interviews to be sufficient. In our setting, saturation was indeed achieved with even fewer interviews; the revealed common and recurring defects can be assumed to be comprehensive.

In setting B, the interviews revealed 3 technical and 1 process-related defect. The technical defects were related to initial values in C++ (2) and overflows (1). The process-related defect were due to interfaces (2) and incomplete specifications (2). Contrary to all other settings, this setting was only to give a first glance as described in the case description. Thus, comprehensiveness was intentionally neglected, but to provide a first glance 3 interviews were sufficient.

In setting C, the interviews revealed 5 technical defects and 6 process-related defects. The technical defects were related to web browser incompatibilities (4), validation of input data (3) and exception handling (2). The most prominent process-related defects were incomplete specification (5) and interface incompatibilities (3). With only 6 interviews, we did not perform a sufficient number of interviews per se. However, the project's size was only 10 persons and effects of saturation were quickly observable. This saturation yields an indication towards comprehensiveness, albeit inconclusive.

In setting D, the interviews revealed 3 technical defects and 7 process-related defects. The top technical defects were related to data dependencies (8), exception handling (4), concurrency (3). The prominent process-related defects were incomplete specifications (7), interface incompatibilities (4) and late involvement of the customers (4). Again, 10 interviews are below the sufficiency baseline, but the project's size was only 18 persons. Once again, saturation yields an indication towards comprehensiveness, albeit inconclusive.

Although the aim of the DELICLA approach was only to elicit and classify technical defects, process-related defects were mentioned by interview partners and were classified as well. Many interview partners stated process-related defects as causes for technical defects, yielding a causal relation between some defects. For instance, in the cyber-physical settings, the inconsistent / incomplete specification was described to lead to incorrect signal ranges and wrong initial values. In the information system domain, the format of user stories as use cases without exceptions was described to lead to untested / incorrect exception handling. We did not believe in advance these causalities or process-related defects to be important at first. However, we later realized their potential for deciding whether to (1) employ defect models for quality assurance to detect or (2) make organizational, methodological or process adjustments to prevent these defects. Many project partners were intrigued about the causalities and estimated the effort to change their processes lower than to employ defect models for some technical defect. In particular, since process improvement by using elicited defects has been described in literature \cite{Kalinowski:2008, Kalinowski:2010}.

An interesting observation was the presence of domain independent defects of technical as well as process-related nature (see Fig.~\ref{fig:defects}). Domain independent technical defects were run time failure causing defects in embedded systems in setting A and B as well as untested exception handling defects in information systems in settings C and D. Moreover, interface incompatibilities and incomplete / inconsistent specifications/requirements  were process-related defects present in all settings.

We therefore conclude so far that our approach is suitable to elicit a broad spectrum of defects which cover the particularities of the envisioned context as well as context-independent defects.

\textbf{RQ 2: Operationalizability.}
In all settings, we were able to derive possible solution proposals for handling each elicited and classified defect. These solution proposals do not necessarily include formal descriptions of defect models, but rather are indicators for operationalization possibilities. However, we or our project partners were able to design tools or methods that have an underlying defect model, and in some cases described next, we were able to operationalize the defects via tools.

Setting A resulted in a testing tool called 8CAGE \cite{Holling:2014}. 8Cage is a lightweight testing tool for Matlab Simulink systems based on defect models. The employed defect models target overflow/underflow, division by zero run time failures as well as signal range and scaling problems.

Setting B yielded an internal testing tool for the testing of the interfaces of the software to the AUTOSAR Runtime Environment (RTE) developed by the project partner. Due to frequent changes in the communication between each electronic control unit, the RTE had to be recreated for each change. Sometimes changes were not implemented leading to unusual failure message and large efforts spent in fault localizations. The internal testing tool can now be run to show these unimplemented changes automatically.

Settings C and D did not result in any tools as of now. However, they yielded requirements to specifically test exception handling functionality in Java systems. The task of the tool is to explicitly throw exceptions at applicable points in the code as to deliberately test the developed exception handling. We currently have collected these requirements and tool development is imminent. In addition, quality standards for SCRUM user story standards as a method of early defect detection have been proposed and partially implemented in setting C and B. These methods include perspective-based reading \cite{Shull2000} of user stories before accepting them and explicit definitions of acceptance criteria including a specification for exception handling.

Overall, the tools and methods developed enable front-loading of quality assurance activities. This allows developers and testers to focus on common and recurring defects in specification and implementation and either makes them aware of the defects or allows the (semi-)automatic detection. Thus, the defect-based perspective may be able to increase the potential to avoid these defects in the future.

We therefore conclude so far that we could elicit defects suitable for operationalization in the chosen context.
	
\textbf{RQ 3: Indirect short-term benefit.}
After presenting the results in the workshop meeting of DELICLA, we asked the responsibles to assess the usefulness of our approach in terms of (1) being aware of the elicited and classified defects, (2) future actions based on the report, and (3) cost-effectiveness. 

In setting A, the responsibles deemed the results satisfying. The defects were mostly known to them, but they were content to have written results in hand for justification towards their management. Using the results, they could convince their management and customers to invest into consulting regarding specific defects. The efficiency of only one hour per interview while leading to sufficiently comprehensive results was perceived positively. They agreed to perform further interviews in the future. However, they remarked that our approach did not reveal many defects previously unknown to them, but were now able to gain an essential understanding of their frequency. They also commented on the difficulty to select distinct projects for the proposed measures in this inter-project setting.

In setting B, the project responsibles only gave us a limited feedback. They stated all defects to be known and saw the advantage in now having a thorough documentation. We did not create defect models or develop operationalizations for them after applying the approach. However, they developed a tool without our involvement based on one reported defect.

In setting C, the project responsibles were surprised how non-intruding and conciliatory our approach is and how professionally it can be handled. They were aware of most of the defects, but not that 20\% of their test cases were already defect-based. The project was already in a late stage when we applied the approach and future actions could not be taken due to the time left. They also perceived the efficiency of one hour per interview partner as positive and described the comprehensiveness of results as given. When discussing further interviews, they questioned the application towards a whole organization as measure to find organization-wide defects with a small number of interviews.

In setting D, the project responsibles were aware of most defects elicited and classified and satisfied with the application of the approach in general. They said the approach ``yields good results with little effort'' and it provides ``a view'' on the defects in the project from ``a different side''. In addition, they stated that ``nothing is missing [from the results] and [results are] diagnostically conclusive''. Concerning the possible solution approaches presented, they ``may not be the way to go'', but ``give a first idea for discussion in project meetings''. Again, further interviews were discussed, but the time required to interview the complete project with more than 50 employees was deemed to much. The project partner rather wanted to use other techniques such as observation or focus groups to minimize time required on their side. However, qualitative interviews were deemed ``a good starting point''.

\vspace{-1.5em}
\subsection{Threats to Validity}
\vspace{-0.8em}
There is a plethora of threats to the validity, let alone those inherent to case study research. To start with, the qualitative nature of the approach as well as the qualitative nature of the evaluation technique rely to a large extent on subjective expert judgment. First and foremost, the approach was applied by the same person evaluating it. The \emph{internal validity} is particularly threatened by the subjectivity in the interviews and especially in their interpretation. Coding used to classify the defects, for example, is an inherently creative task. However, our aim was to explore the potential of such qualitative methods, to reveal subjective opinions by the study participants, and to elaborate -- despite the inherent threats -- the suitability of the chosen approach. 

The \emph{construct validity} is threatened in two ways. First, the research questions were answered via qualitative methods only and we cannot guarantee that we could fully answer the questions based on our data. We compensated the threat, especially for research question 1, by taking an external baseline as an orientation. Second, we cannot guarantee that we have a sufficient number of interviews to reliably decide on the completeness of the data to elaborate comprehensive defects. We compensated this threat by applying the principles of Grounded Theory where we explicitly considered a saturation of the answers if no new codes arose. Also, we believe the number of interviews to be less important than the coverage of roles within (different) teams and superordinate roles. This is hinted at by setting B, C and D in particular. 

Finally, the \emph{external validity} is threatened by the nature of the approach and the evaluation as well. Our intention was, however, not to generalize from our findings but to evaluate the extent to which our approach is suitable to cover the particularities of contexts whereby the results hold specifically for those contexts. Yet, by comparing the results with an external baseline, we could determine context-independent defects which potential for generalization.

\vspace{-1.5em}
\section{Conclusion}
\label{sec:conclusion}
\vspace{-1em}
We have evaluated our DELICLA approach to elicit and classify defects for their eventual description and operationalization as defect models. DELICLA is entirely based on existing elicitation and analysis approaches \cite{Gubrium:2001}. Our approach 
uses a qualitative explorative method with personal interviews as elicitation and grounded theory as classification technique. 
We have evaluated the approach in a field study with four different companies. The chosen settings varied in their domains. Using our approach, we were able to elicit and classify defects having both a context-dependent and -independent relevance while providing indicators to the extent to which they relate to existing evidence. The approach was applicable in different contexts/domains due to the employment of qualitative approaches. We could use the elaborated defects to derive requirements for their (semi-) automatic detection by tools and create possible solution proposals in all settings. The feedback given to us by project partner executives was positive yielding ``informative results with little effort''. Thus, the results strengthen our confidence that the studies are representative and the approach is suitable to elicit context-specific defects without being too specific for a context. However, we are lacking a study with negative results.



Using DELICLA, we have gained an insight of existing defects in different contexts and domains. For some defects, we have built and operationalized respective fault models. Moreover, for some defects we were able to find cause effect relationships to other defects. This yields the question about the ability of the method to extensively find cause effect relationships for all defects. To answer this question, further work and maybe a higher level portrayal of defects as provided by the methodology of Kalinowski et. al \cite{Kalinowski:2010} is required.  





\vspace{-1em}
\bibliographystyle{lncs}
\bibliography{literature}
\vspace{-1em}

\end{document}